\documentclass[a4paper]{jpconf}
\newcommand{\ks} {{\bf k}}
\newcommand{\vs} {{\bf v}}
\newcommand{\esm} {{\bf e}}

\newcommand{\ec} {{\bf E}}
\newcommand{\pc} {{\bf P}}
\newcommand{\jc} {{\bf J}}
\usepackage{graphicx}
\begin{document}
\title{Blueshift of high-order harmonic generation in crystalline silicon subjected to intense femtosecond near-infrared laser pulse}

\author{Boyan Obreshkov$^1$, Tzveta Apostolova$^{1,2}$}

\address{$^1$ Institute for Nuclear Research and Nuclear Energy, Bulgarian Academy of Sciences, Tsarigradsko chausse 72, 1784 Sofia, Bulgaria}
\address{$^2$ Institute for Advanced Physical Studies,
New Bulgarian University, 1618 Sofia, Bulgaria}

\ead{}

\begin{abstract}

We present the generation of high order harmonics in crystalline silicon subjected to intense near-infrared 30fs laser pulse. The harmonic spectrum extends from the near infrared to the extreme ultraviolet spectral region. Depending on the pulsed laser intensity,
we distinguish two regimes of harmonic generation:
(i) perturbative regime: electron-hole pairs born during each half-cycle of the laser pulse via multiphoton and tunnel transitions are accelerated in the laser electric field and gain kinetic energy; the electron-hole pairs then recombine in the ground state by emitting a single high-energy photon. The resultant high harmonic spectrum consists of sharp peaks at odd harmonic orders.
(ii) non-perturbative regime: the intensity of the harmonics increases, their spectral width broadens and the position of harmonics shifts to shorter wavelengths. The blueshift of high harmonics in silicon are independent on the harmonic order which may be helpful in the design of continuously tunable  XUV sources.
\end{abstract}

\section{Introduction}

High harmonic generation (HHG) is a nonlinear optical process which occurs when a target (atomic gas, plasma, amorphous or crystalline solid) is irradiated by an intense laser beam. Subsequently the photoexcited state emits radiation in the form of high harmonics of the driving laser frequency.  HHG in atomic gases is usually interpreted within a semi-classical three-step model \cite{Krause1992,Corkum1993} including photoionization, acceleration of the freed electron by the laser electric field and subsequent radiative re-combination of the electron with its parent ion resulting in emission of single high-energy photon.  A typical HHG spectrum in the gas-phase consists of odd order harmonics with a plateau region and sharp cutoff at $\hbar \omega_{{\rm max}}=I_p+3.17 U_p$, where $I_p$ is the atomic ionization potential and $U_p=e^2 E^2/4 m \omega_L^2$ is the ponderomotive energy; here $e$ is the electronic charge, $m$ is the electron mass, $E$ is the peak electric field strength and $\omega_L$ is the driving laser frequency.  

More recently HHG has been observed in bulk ZnO crystals subjected to intense mid-infrared laser pulses \cite{Ghimire2011}.  The measured HHG spectrum displays features similar to the gas phase with high harmonic orders extending into the UV region. In contrast to gas phase harmonics, the cutoff energy for solid-state HHG was found to scale linearly with the peak field strength. Also the measured yield of high harmonics differs substantially from the prediction of perturbative nonlinear optics.  These observations suggest that strong-field  driven electron dynamics in solids is qualitatively different from the prediction of the semi-classical three-step model applicable to the gas phase. Subsequently, HHG was demonstrated in wide-band gap solids \cite{Schubert2014,Luu2015,Ndabashimiye,You2017} and in graphene \cite{Yoshikawa2017}. On this basis solid-state HHG has promising potential in development of compact ultrafast and coherent XUV sources (cf. Ref. \cite{Garg2018}).  

In this paper we present numerical results of HHG in bulk silicon driven by linearly polarized 30fs laser pulse of 800 nm wavelength: Sec. 2 presents the time evolution of the electric field of the transmitted pulse and the characteristics of  HHG spectra in crystalline silicon as a function of the peak laser intensity, Sec. 3 includes our main conclusion.

\section{Numerical results and discussion}

Details of the theoretical approach and methodology can be found in Refs. \cite{DRM2018,OQEL2018,Apsusc2020}. In brief, we solve numerically the time-dependent Schr\"{o}dinger equation for valence electrons subjected to a linearly polarized intense ultrashort laser pulse with near-infrared wavelength 800 nm. The time-dependent electric field of the transmitted pulse is 
presented as $\ec=\ec_{{\rm ext}}+\ec_{{\rm ind}}$; the applied laser electric field 
is parameterized by a temporary Gaussian function 
\begin{equation}
\ec_{{\rm ext}}(t)=\esm  e^{-\ln(4) (t-t_0)^2/\tau_L^2} F \cos \omega_L t
\end{equation}
where $\esm$ is the laser polarization vector, $\omega_L$ is the laser oscillation frequency (corresponding to photon energy $\hbar \omega_L=1.55$ eV),
$\tau_L=$ 30 fs is the pulse length, $t_0$ specifies the position of the pulse peak, $F=I^{1/2}$ is the electric field strength and $I$ is the peak laser intensity.  The induced electric field $\ec_{{\rm ind}}=-4 \pi \pc$ is a result of the polarization of the solid, here the macroscopic polarization $\pc(t)=\int^t dt' \jc(t')$ is expressed in terms of the transient photo-current $\jc=tr[\rho \vs]$, where $\rho$ is the one-electron density matrix and $\vs$ is the velocity operator. 

The static band structure of silicon is obtained from empirical pseudo-potential method. The direct bandgap  energy (3.2 eV) associated with the $\Gamma_{25'} \rightarrow \Gamma_{15}$ interband transition specifies the threshold for electron-hole pair excitation. To calculate the transient photocurrent $\jc$, we sample the  Brillouin zone by a Monte Carlo method using 5000 quasi-randomly $\ks$-points generated from three-dimensional Sobol sequence in a cube of edge length $4 \pi/a_0$, where $a_0=$ 5.43 \AA ~is the bulk lattice constant of Si;  4 valence and 16 initially unoccupied conduction bands are included in the expansion of the time-dependent wave-functions over static Bloch orbitals. The wave-functions of valence electrons were propagated forward in time for small equidistant time steps of $\delta t=$ 0.7 attoseconds. The spectrum  of high harmonic generation inside the bulk is obtained from the Fourier transformation of the component of the photocurrent  onto the laser polarization direction 
\begin{equation}
I(\omega)= \left| \int dt e^{i \omega t} \jc(t) \cdot \esm \right|
\end{equation}

\begin{figure}[h]
\centering\includegraphics[width=1\linewidth]{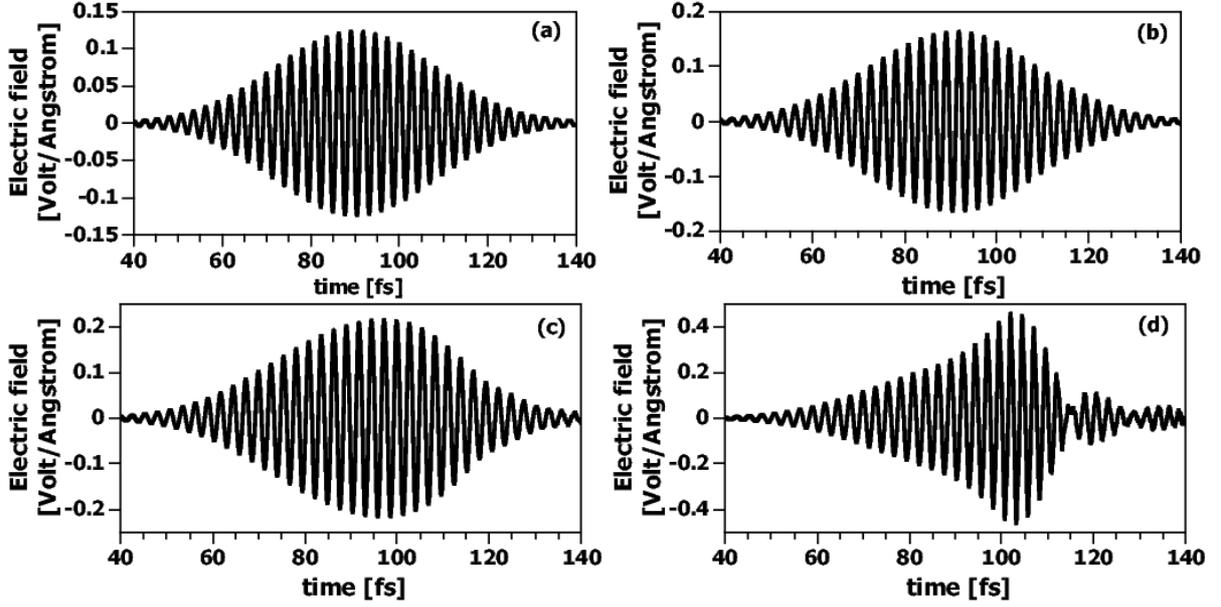}
\caption{{(a-c) Time evolution of the pulsed electric field (in V/\AA) transmitted in bulk silicon. The laser intensity at the pulse peak is 3 $\times $ 10$^{13}$ W/cm$^2$ in (a), 5 $\times $ 10$^{13}$ W/cm$^2$ in (b), 7 $\times $ 10$^{13}$ W/cm$^2$ in (c) and 9 $\times $ 10$^{13}$ W/cm$^2$ in (d). The laser is linearly polarized along the [001] direction, the laser wavelength is 800 nm and the pulse duration is 30 fs. }
}
\label{fig:efield}
\end{figure}

\begin{figure}[h]
\centering\includegraphics[width=1\linewidth]{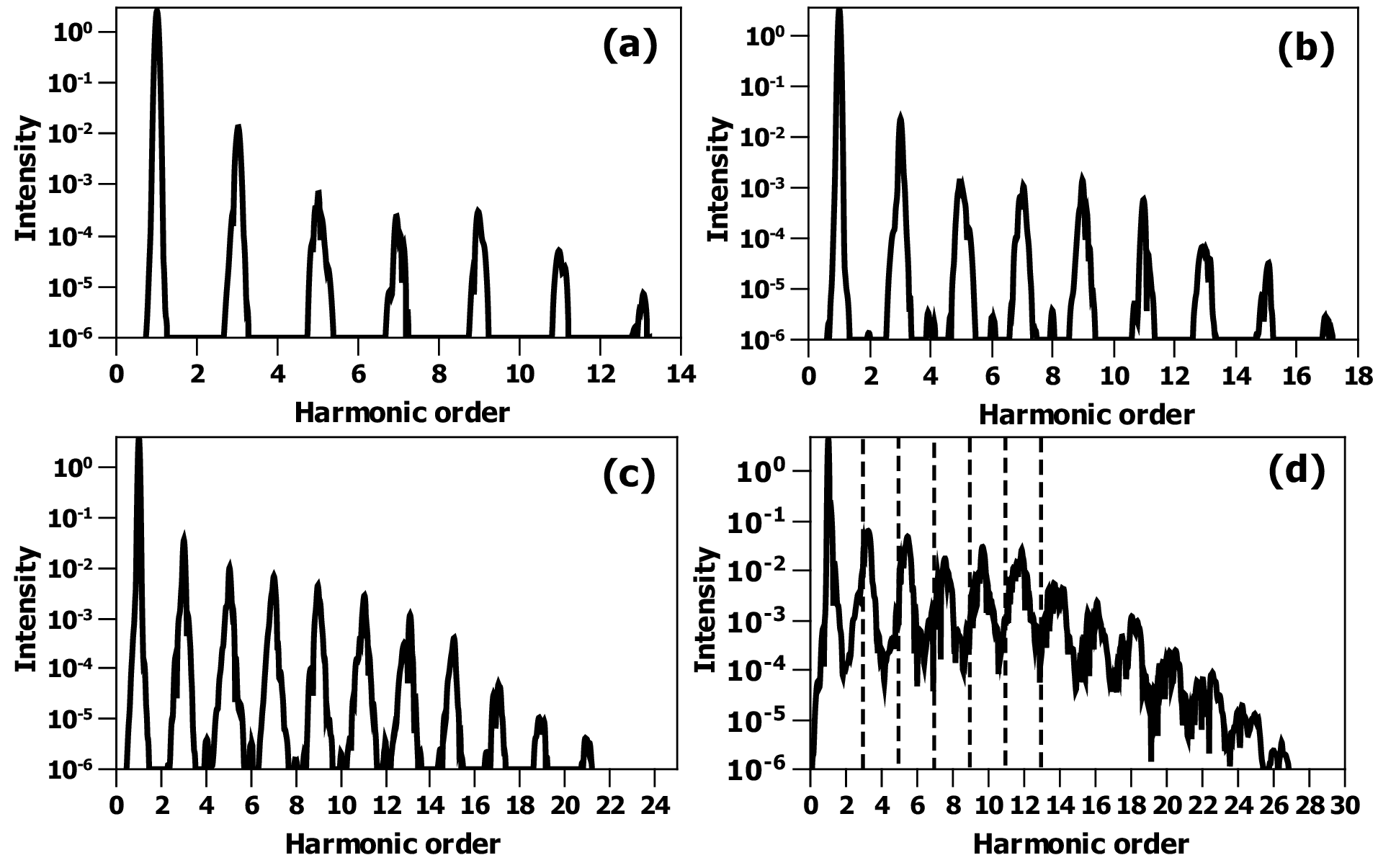}
\caption{ HHG spectra is bulk silicon as a function of the peak laser intensity (a) $I=$ 3 $\times 10^{13}$ W/cm$^2$, (b) $I=$ 5 $\times 10^{13}$ W/cm$^2$, (c) $I=$ 7 $\times 10^{13}$ W/cm$^2$ and (d) $I=$ 9 $\times 10^{13}$ W/cm$^2$. The laser is linearly polarized along the [001] direction, the laser wavelength is 800 nm and the pulse duration is 30 fs. }
\label{fig:hhg}
\end{figure}

The time evolution of the electric field of the transmitted laser pulse in bulk silicon 
is shown in Fig.\ref{fig:efield}. For the relatively low laser intensity, $I=3 \times 10^{13}$ W/cm$^2$ in  Fig.\ref{fig:efield}a, the pulse exhibits temporary Gaussian profile with field strength $\ec=\ec_{{\rm ext}}/12$ and $\epsilon \approx 12$ is the static bulk dielectric constant of Si. 
The associated HHG spectrum shown in Fig. \ref{fig:hhg}a consists of clean odd order harmonic peaks, i.e. close to zero crossing of the oscillating electric field  electron-hole pairs recombine and XUV photons are emitted during each half-cycle. Though photoionization involves perturbative three-photon transition across the direct bandgap in the low-intensity regime,  non-perturbative effects in HHG are exhibited as the spectral intensities of the 5th to 9th harmonics are of comparable magnitude and comprise a plateau region. This effect can be traced to the accelerated motion  of charge carriers in their respective bands. For the increased peak intensity, shown in  Fig.\ref{fig:efield}b,c the peak of the transmitted pulse undergoes a progressive time delay relative to the peak of the applied laser. For $I=7 \times 10^{13}$ W/cm$^2$, Fig.\ref{fig:efield}c the trailing edge of the pulse becomes steeper as a result of increasing population in the conduction band. Both the harmonic yield and the cutoff energy for HHG increase moderately with the increase of intensity, cf. Fig. \ref{fig:hhg}b,c. 

For the highest peak intensity shown in  Fig.\ref{fig:efield}d, $I=9 \times 10^{13}$ W/cm$^2$, the non-linear response of electrons becomes prominent: the temporal profile of the pulse is strongly distorted, the peak intensity is pushed to the back of the pulse where a steep edge is formed and frequency up-chirp is exhibited. This regime corresponds to high level of electronic excitation: each Si atom absorbs 1.2 eV energy (i.e. each atom absorbs nearly one laser photon). The harmonic peaks, shown in Fig.\ref{fig:hhg}d, superimpose onto a continuous background, the spectral width of individual harmonics broadens and their central wavelength undergoes a blue shift that covers the spacing between adjacent orders. Noticeably the position of the fundamental harmonic is unchanged and the shift of the position of harmonic peaks $\delta \omega_q \approx  0.1 \omega_L$ is independent on the harmonic order $q$, i.e. harmonics are mainly emitted at the back of the pulse when the driving frequency is up-chirped.

\section{Conclusion}

In summary we presented calculation of HHG spectrum in bulk silicon induced by intense 30fs near-infrared laser pulse. The result describes HHG in a wide range of peak laser  intensities. In the low intensity regime $I \le 5 \times 10^{13}$ W/cm$^2$, the temporal profile of the transmitted pulse is weakly distorted by the laser-matter interaction and exhibits relatively high degree of temporal coherence. In this regime, HHG spectrum in silicon consists of clean odd-order harmonics with plateau region and sharp cutoff.  The temporal coherence of the transmitted pulse deteriorates with the increased laser intensity $I \ge 7 \times 10^{13}$ W/cm$^2$, the peak of the pulse undergoes progressive time delay relative to the applied laser and becomes subject to self-steepening. In this regime, the up-chirp in the back part of the pulse results in blueshift of high-order harmonic generation in silicon. 

\section*{Acknowledgment}

The material is based upon work supported by the Air Force Office of Scientific Research under award number FA9550-19-1-7003. Support from the Bulgarian National Science Fund under contracts No. 08-17 (B.O) and No. KP-06-KOST (T.A) is acknowledged.

\end{document}